\documentclass[11pt,a4paper]{article}

\usepackage[T1]{fontenc}
\usepackage[utf8]{inputenc}

\usepackage[top=2.5cm,bottom=2.5cm,left=2.8cm,right=2.8cm]{geometry}
\usepackage{setspace}
\usepackage{amsmath,amssymb,amsthm,mathtools,cancel}

\usepackage{booktabs,array,tabularx,multirow}
\usepackage{float}

\usepackage{enumitem}

\usepackage[hidelinks]{hyperref}

\theoremstyle{plain}
\newtheorem{theorem}{Theorem}[section]
\newtheorem{lemma}[theorem]{Lemma}
\newtheorem{proposition}[theorem]{Proposition}

\theoremstyle{definition}
\newtheorem{definition}[theorem]{Definition}

\numberwithin{equation}{section}

\DeclareMathOperator{\sgn}{sign}

\begin{document}
	
	\title{\bfseries Fixed Capital, the Cost Criterion, and the Falling Rate of Profit}
	\author{}
	\author{Lyu Jiyuan\thanks{School of Economics and Management, Beijing University of Technology. Email: sergeylyu@163.com}}
	
	\date{}
	
	\maketitle
	
	\begin{abstract}
		
		Existing studies that introduce the concept of fixed capital into the Okishio Theorem show that, under an unchanged real wage, cost-reducing technical progress still necessarily raises the equilibrium rate of profit. This paper argues that this conclusion depends on how cost is defined in fixed capital models. In the original model of Okishio (1961), cost corresponds to what is termed operating cost in accounting, which excludes any form of profit; by contrast, the classical fixed capital frameworks adopted by Roemer (1979) and Woods (1985) cannot mathematically separate depreciation from profit, so that cost comparison in fact corresponds to capitalized cost inclusive of profit. This paper employs the annuity method to treat fixed capital, thereby separating depreciation from profit, and grounds the capitalist's technology choice on the basis of operating cost. The results show that two distinct thresholds exist in the fixed capital model: a profit-rate threshold that determines the direction of change in the rate of profit after the diffusion of the new technology, and an operating-cost threshold that determines whether the capitalist will adopt the new technology. When the lifetime of fixed capital exceeds one production period, the two thresholds are strictly separated, giving rise to a nonempty interval. Within this interval, the new technology lowers operating cost and is adopted by capitalists, yet the new equilibrium rate of profit falls below its initial level. The pure circulating capital model corresponds to the special case in which the two thresholds coincide, so that the Okishio Theorem can be regarded as a special case of the results obtained in this paper.
		
		\smallskip
		\noindent
		\textbf{Keywords:}
		falling rate of profit; Okishio Theorem; fixed capital; cost criterion; rate of profit
	\end{abstract}
	
	\section{Introduction}
	
	The law of the tendency of the general rate of profit to fall is one of the most controversial issues in Marxian political economy. In Volume~III of \textit{Capital}, Marx argues that, as capital accumulation proceeds and the organic composition of capital rises steadily, the general rate of profit possesses an inherent tendency to fall. This proposition has long served both as an important theoretical foundation of Marxian economics and as a focus of debate surrounding the labor theory of value and the laws of capitalist development.
	
	The most influential formal challenge to this proposition is the Okishio Theorem, put forward by Okishio (1961). Within a multisectoral input--output framework, Okishio proved that if capitalists adopt a new technology that reduces cost, and if the real wage remains unchanged, then the general rate of profit in the new equilibrium must be higher than its original level. The theorem's cost criterion is consistent with Marx's fundamental assumption about capitalist behavior, namely that capitalists choose technology on the basis of cost rather than labor productivity. Okishio distinguished between a productivity criterion and a cost criterion for technology choice: the former requires that the direct and indirect labor expended per unit of output decrease, whereas the latter merely requires that the new technology reduce unit cost at prevailing prices and wages. In a capitalist economy, capitalists follow the latter criterion. Precisely for this reason, the Okishio Theorem is generally regarded as a challenge to the law of the falling rate of profit raised from within Marx's own analytical framework.
	
	Discussions surrounding the Okishio Theorem have continued to the present day. Shaikh (1978) questioned its methodological premises from the perspective of disequilibrium competition; Kliman (1996) and Nakatani (2005) re-examined the mechanism of profit-rate movements within the framework of the Temporal Single-System Interpretation (TSSI); Laibman (1982), Luo Zhen (2010), Xue Yufeng (2012), and Li Bangxi et al.\ (2016) also discussed, from different angles, the relationship between the Okishio Theorem and the law of the falling rate of profit. Yang Shuaihong and Zhu Andong (2021) relaxed the constant-real-wage assumption within a fixed capital framework and likewise obtained results in which the rate of profit can fall. However, most of these studies alter the basic analytical framework of the Okishio Theorem, and therefore do not directly unsettle its conclusion within the comparative-static equilibrium model.
	
	Another line of research consists in introducing fixed capital into the model while preserving Okishio's method of analysis. Roemer (1979) and Woods (1985) generalized the Okishio Theorem within, respectively, a von~Neumann fixed capital model and a Sraffian joint-production model, both demonstrating that cost-reducing technical progress still necessarily raises the general rate of profit. Salvadori (1981) constructed a counterexample within a joint-production framework in which cost falls while the rate of profit falls, but Fujimoto (1983) pointed out that this result depends on the cross-price effects generated by joint production, rather than on fixed capital as such. Hence the existing literature generally holds that, so long as the basic assumptions of the Okishio Theorem are maintained, fixed capital itself does not alter its core conclusion.
	
	This paper contends that this conclusion rests on a premise that has received relatively little discussion---namely, how cost ought to be defined in the presence of fixed capital.
	
	In the original model of Okishio (1961), the price equation is written as
	\[
	q_i=(1+r)\left(\sum_j a_{ij}q_j+\tau_i\right),
	\]
	and the cost comparison concerns the intermediate inputs and direct labor inside the parentheses, with the profit-rate factor $(1+r)$ being excluded from cost. Cost in Okishio's sense therefore contains no form of return on capital. This treatment is consistent with Marx's definition of cost price and with the distinction drawn in modern accounting standards between operating cost and finance charges. In a pure circulating capital model, since all means of production are entirely used up within a single production period and no depreciation problem arises, this definition creates no further difficulty.
	
	Once fixed capital enters the model, the situation changes. The user cost of fixed capital can be decomposed into two components---depreciation and the return on capital (profit)---which differ in their economic meaning. If one follows Okishio's original cost criterion, depreciation should be included in cost, whereas the return on capital should remain excluded. However, in Roemer's (1979) von~Neumann framework, the value of fixed capital is endogenously determined by equilibrium prices, so that depreciation cannot be defined independently of the rate of profit; in Woods's (1985) joint-production framework, the time-weighting of fixed capital directly incorporates the profit-rate factor, which likewise makes it impossible to construct a depreciation cost independent of the rate of profit. As a result, both generalizations in fact adopt a cost concept that includes the return on capital, which is inconsistent with the cost criterion of Okishio (1961).
	
	This paper re-examines the issue. In order to restore the distinction between depreciation and the return on capital, the annuity method is adopted to treat fixed capital, so that the user cost of fixed capital can be expressed as the sum of depreciation and the return on capital; accordingly, the cost on which capitalists base their technology choice is defined as the sum of intermediate inputs, direct wages, and depreciation, excluding any form of interest. Within this framework, two distinct thresholds emerge in the fixed capital model: a smaller one that determines whether capitalists are willing to adopt the new technology according to the cost criterion, and a larger one that determines whether the rate of profit falls after the adoption of the new technology. When the lifetime of fixed capital exceeds one production period, the two thresholds are strictly separated, giving rise to a nonempty region. Within this region, the new technology reduces the operating cost faced by capitalists and is therefore voluntarily adopted, leading to a lower uniform rate of profit. This result depends neither on joint production nor on wage changes or disequilibrium adjustment, but is obtained within the comparative-static framework of Okishio (1961). In the limiting case of pure circulating capital, the two thresholds coincide and the results of this paper reduce to the Okishio Theorem.
	
	The remainder of the paper is organized as follows. Section~2 sets up the price system with fixed capital and discusses the two cost concepts---operating cost and capitalized cost. Section~3 analyzes single-sector technical change and establishes the existence and properties of the profit-rate threshold and the cost threshold. Section~4 provides numerical examples. Section~5 discusses the relationship between the results of this paper and those of Roemer (1979), Woods (1985), and Salvadori (1981), and analyzes their economic implications. The final section concludes.

	\section{The Model}
	\label{sec:model}
	
	\subsection{The Structure of Production}
	
	Consider a closed economy consisting of one capital-good sector and \(n\) consumption-good sectors.
	Sector \(0\) produces the capital good, and sectors \(1,\dots,n\) produce consumption goods.
	
	Intermediate inputs among the consumption-good sectors are represented by the input--output matrix
	\[
	A=(a_{ij})_{n\times n},
	\]
	where \(a_{ij}\) denotes the input of consumption good \(i\) required to produce one unit of consumption good \(j\).
	The consumption-good inputs required to produce one unit of the capital good are denoted by
	\[
	\mathbf a_0=(a_{01},\dots,a_{0n})^\top.
	\]
	
	Direct labor inputs in each sector are denoted respectively by
	\[
	\mathbf l_c=(l_1,\dots,l_n)^\top,\qquad l_0,
	\]
	and fixed capital stocks are denoted respectively by
	\[
	\mathbf k_c=(k_1,\dots,k_n)^\top,\qquad k_0,
	\]
	where all fixed capital is measured in physical units of the capital good.
	
	Fixed capital is accounted for using the annuity method. Let the depreciation rate be
	\[
	\delta\in(0,1],
	\]
	let the uniform rate of profit be \(r\), and denote
	\[
	R\equiv1+r,\qquad
	\sigma\equiv r+\delta=R-1+\delta.
	\]
	Thus the annualized user cost of one unit of fixed capital is \(\sigma p_0\).
	
	Okishio (1961) assumes that wages are paid in advance, and the same treatment is followed here: wages are advanced at the beginning of the production period and, together with intermediate inputs, constitute the circulating capital advanced.
	
	\subsection{Price Equations and the Wage Constraint}
	
	The capital good is taken as num\'{e}raire:
	\[
	p_0\equiv1.
	\]
	Prices of consumption goods and of the capital good satisfy, respectively,
	\begin{align}
		\mathbf p^\top
		&=
		R\bigl(\mathbf p^\top A+w\mathbf l_c^\top\bigr)
		+\sigma\mathbf k_c^\top,
		\label{eq:price_c}\\
		1
		&=
		R\bigl(\mathbf p^\top\mathbf a_0+w l_0\bigr)
		+\sigma k_0.
		\label{eq:price_0}
	\end{align}
	
	Equation~\eqref{eq:price_c} gives the price equations for the consumption-good sectors,
	and equation~\eqref{eq:price_0} corresponds to the capital-good sector.
	Since fixed capital is treated via the annuity method, its user cost consists of both depreciation and the return on capital, and is therefore expressed as \(\sigma=r+\delta\).
	
	The real wage basket
	\[
	\mathbf b\in\mathbb R_+^n,\qquad
	\mathbf b\neq\mathbf0,
	\]
	is exogenously given and held constant. The money wage is determined by
	\begin{equation}
		\mathbf b^\top\mathbf p=w.
		\label{eq:wage}
	\end{equation}
	
	Throughout the paper the following three basic assumptions are maintained.
	
	\begin{enumerate}[label=(H\arabic*),nosep,leftmargin=*]
		\item The input--output matrix satisfies
		\[
		\rho(A)<1,
		\]
		i.e., the consumption-good production system is productive.
		
		\item
		\[
		\mathbf l_c,\,
		\mathbf k_c,\,
		\mathbf a_0,\,
		l_0,\,
		k_0,\,
		\mathbf b
		\]
		are all nonnegative vectors (or scalars) and each is nonzero.
		
		\item The economic system is irreducible; that is, there exists no nonempty proper subset
		\[
		J\subset\{0,1,\dots,n\}
		\]
		such that all sectors depend only on inputs from within \(J\) to carry out production.
	\end{enumerate}
	
	\subsection{Effective Total Labor and Profit-Rate Comparison}
	
	Define the augmented matrix
	\[
	\tilde{\mathbf M}_\delta(R)
	=
	\begin{pmatrix}
		\sigma k_0
		&
		R\mathbf a_0^\top
		\\
		\sigma\mathbf k_c
		&
		RA^\top
	\end{pmatrix},
	\]
	and
	\[
	\tilde{\mathbf l}
	=
	\begin{pmatrix}
		l_0\\
		\mathbf l_c
	\end{pmatrix},
	\qquad
	\tilde{\mathbf b}
	=
	\begin{pmatrix}
		0\\
		\mathbf b
	\end{pmatrix},
	\qquad
	\tilde{\mathbf p}
	=
	\begin{pmatrix}
		1\\
		\mathbf p
	\end{pmatrix}.
	\]
	
	The price system~\eqref{eq:price_c}--\eqref{eq:wage} can be written as
	\begin{equation}
		(\mathbf I-\tilde{\mathbf M}_\delta(R))
		\tilde{\mathbf p}
		=
		Rw\tilde{\mathbf l},
		\qquad
		\tilde{\mathbf b}^\top\tilde{\mathbf p}
		=
		w.
		\label{eq:augmented}
	\end{equation}
	
	Define the effective total labor function
	\begin{equation}
		V(R)
		=
		\tilde{\mathbf b}^\top
		(\mathbf I-\tilde{\mathbf M}_\delta(R))^{-1}
		\tilde{\mathbf l}.
		\label{eq:V_def}
	\end{equation}
	
	From~\eqref{eq:augmented} one immediately obtains
	\begin{equation}
		RV(R)=1.
		\label{eq:RV_identity}
	\end{equation}
	
	Under Assumption~(H3), by the theory of irreducible nonnegative matrices and M-matrices,
	\(V(R)\) is strictly increasing on its domain.
	(See Appendix~A.1 for the proof.)
	
	\begin{definition}
		Given technological parameters, if a triple
		\[
		(r,\mathbf p,w)
		\]
		satisfies
		\[
		r>0,\qquad
		\mathbf p>\mathbf0,\qquad
		w>0,
		\]
		and simultaneously satisfies the price equations~\eqref{eq:price_c}--\eqref{eq:price_0}
		and the wage constraint~\eqref{eq:wage},
		then it is called an \emph{equilibrium} of the economy.
	\end{definition}
	
	\begin{proposition}[Existence and Uniqueness of Equilibrium]
		\label{prop:existence}
		Under Assumptions~(H1)--(H3), if
		\[
		V(1)<1,
		\]
		then an equilibrium exists and is unique.
	\end{proposition}
	
	\begin{proof}
		See Appendix~A.1.
	\end{proof}
	
	Denote the profit-rate factor corresponding to the initial equilibrium by
	\[
	R_0=1+r_0,
	\qquad
	V_0=V(R_0).
	\]
	
	Using block matrix inversion, \(V_0\) can be expressed as
	\begin{equation}
		V_0
		=
		\mathcal L
		+
		\frac{\sigma_0\ell_0\mathcal K}
		{\Delta_\delta},
		\label{eq:V_decomp}
	\end{equation}
	where
	\begin{align}
		\mathcal L
		&=
		\mathbf b^\top
		(\mathbf I-R_0A^\top)^{-1}
		\mathbf l_c,
		\label{eq:L_def}
		\\
		\mathcal K
		&=
		\mathbf b^\top
		(\mathbf I-R_0A^\top)^{-1}
		\mathbf k_c,
		\label{eq:K_def}
		\\
		\ell_0
		&=
		l_0
		+
		R_0
		\mathbf a_0^\top
		(\mathbf I-R_0A^\top)^{-1}
		\mathbf l_c,
		\label{eq:ell0_def}
		\\
		\Delta_\delta
		&=
		1
		-
		\sigma_0k_0
		-
		\sigma_0R_0
		\mathbf a_0^\top
		(\mathbf I-R_0A^\top)^{-1}
		\mathbf k_c.
		\label{eq:Delta_def}
	\end{align}
	
	The derivation by block inversion is given in Appendix~A.2.
	
	\begin{lemma}
		\label{lemma:Delta}
		At the initial equilibrium,
		\begin{equation}
			\Delta_\delta
			=
			R_0w_0\ell_0.
			\label{eq:Delta_identity}
		\end{equation}
	\end{lemma}
	
	\begin{proof}
		See Appendix~A.2.
	\end{proof}
	
	Let \(r_0\) and \(r^{**}\) denote, respectively, the rate of profit in the initial equilibrium and in the equilibrium after the adoption of the new technology, and let the corresponding effective total labor function for the new technology be denoted by \(V^*(R)\).
	
	\begin{lemma}[Profit-rate comparison]
		\label{lemma:equivalence}
		
		The following equivalences hold:
		\[
		r^{**}<r_0
		\Longleftrightarrow
		V^*(R_0)>V_0,
		\]
		\[
		r^{**}>r_0
		\Longleftrightarrow
		V^*(R_0)<V_0,
		\]
		\[
		r^{**}=r_0
		\Longleftrightarrow
		V^*(R_0)=V_0.
		\]
		
	\end{lemma}
	
	\begin{proof}
		See Appendix~A.3.
	\end{proof}
	
	\subsection{Two Cost Concepts}
	
	Once fixed capital is introduced, the cost on which technology choice is based needs to be redefined. The key difference lies in the fact that the user cost of fixed capital can be decomposed into depreciation and the return on capital, and whether both enter cost simultaneously determines different cost criteria.
	
	Modern accounting standards draw a clear distinction in this regard. According to IAS~2 (Inventories) and IAS~23 (Borrowing Costs), depreciation belongs to production cost, whereas interest on borrowings is in principle classified as a finance charge and is included in the cost of an asset only when capitalization conditions are met, subsequently being transferred to operating cost via depreciation. Interest on borrowings is not identical to interest in Marx's sense---the former corresponds to the cost of debt capital, the latter to the return on total advanced capital---but neither belongs to operating cost in the usual sense. Accounting standards thus provide two different cost measures.
	
	The first is \emph{operating cost}. Unit cost includes only intermediate inputs, direct wages, and depreciation of fixed capital:
	\begin{equation}
		C_j^{\mathrm{op}}
		=
		\sum_i a_{ij}p_i
		+
		w_0l_j
		+
		\delta k_j.
		\label{eq:op_cost}
	\end{equation}
	
	The second is \emph{capitalized cost}. If the return on capital is treated as part of the cost of the asset, then unit cost is written as
	\begin{equation}
		C_j^{\mathrm{cap}}
		=
		\sum_i a_{ij}p_i
		+
		R_0w_0l_j
		+
		\sigma_0k_j,
		\label{eq:cap_cost}
	\end{equation}
	where
	\(
	R_0=1+r_0
	\)
	and
	\(
	\sigma_0=r_0+\delta
	\).
	Compared with operating cost, capitalized cost includes not only the interest on fixed capital but also the return on capital advanced as wages.
	
	This distinction is consistent with the discussion of cost price in \textit{Capital}. In Chapter~1 of Volume~III, Marx defines cost price as follows: ``That part of the value of the commodity, then, which makes good the price of the means of production used up and the price of the labour-power employed, only makes good what the commodity costs the capitalist himself, and for him, therefore, represents the cost price of the commodity\dots\ If we call cost price \(k\), the formula \(W = c + v + m\) is converted into the formula \(W = k + m\), or commodity value = cost price + surplus-value,'' i.e., \(c+v\). ``The wear and tear of the instruments of labour'' (including depreciation) is explicitly listed as a component of cost price, while surplus-value \(m\) is excluded from cost price. In Chapter~23 of Volume~III, Marx further points out that interest ``appears originally, is originally, and remains in fact nothing but a part of profit, i.e., surplus-value.'' Hence depreciation belongs to cost price, while interest belongs to the division of surplus-value rather than to cost. Depreciation and interest are thus of different economic character.
	
	Okishio (1961) adopted precisely the operating-cost criterion. In his circulating capital model,
	\[
	q_i
	=
	(1+r)
	\left(
	\sum_j a_{ij}q_j+\tau_i
	\right),
	\]
	the cost comparison concerns the intermediate inputs and direct labor inside the parentheses, and the profit-rate factor \((1+r)\) is excluded from the comparison. Since circulating capital is entirely used up within one production period, the concept of depreciation does not arise, and operating cost reduces to intermediate inputs plus direct labor, which indicates that Okishio's cost criterion does not include any form of interest. Since interest on circulating capital is excluded from cost in Okishio's (1961) model, why should interest on fixed capital be included in cost when fixed capital is introduced into the Okishio Theorem? Roemer's (1979) von~Neumann framework endogenizes the value of machines within the price system, so that depreciation of fixed capital cannot be defined independently of the rate of profit; Woods's (1985) joint-production model uses time weights containing \((1+r)\) to keep the reduced system consistent with the rate of profit of the original system. Neither model can mathematically separate depreciation from the return on capital, and hence their cost comparisons in fact correspond to capitalized cost rather than the operating cost adopted by Okishio (1961).
	
	Because Okishio's (1961) treatment of cost is consistent with the accounting treatment of operating cost, and taking into account the definition of cost in \textit{Capital}, this paper adopts operating cost as the capitalist's technology-choice criterion. Using the annuity method, the user cost of fixed capital is expressed as
	\[
	\sigma k
	=
	\delta k
	+
	rk,
	\]
	which allows depreciation and the return on capital to be treated separately. The capitalist compares intermediate inputs, direct wages, and depreciation, while the return on capital only determines the rate of profit within the equilibrium price system and does not enter the individual's cost comparison.

	\section{Technical Change and Main Results}
	\label{sec:tech}
	
	Consider single-sector labor-saving technical change. Suppose innovation takes place in sector \(j\) (either the capital-good sector or any one consumption-good sector), with all other sectors' technologies unchanged. The labor input and fixed capital stock in the innovating sector change respectively to
	\begin{equation}
		l^{*}=\lambda l,\qquad
		k^{*}=\kappa k,
		\label{eq:tech}
	\end{equation}
	where
	\[
	0<\lambda<1,\qquad
	\kappa\ge1.
	\]
	The parameter \(\lambda\) measures the degree of labor saving, and \(\kappa\) measures the degree of capital deepening. The real wage basket \(\mathbf b\) remains unchanged before and after the technical change.
	
	To highlight the innovating sector \(j\), its labor input and fixed capital are denoted with capital letters, while those of other sectors continue to be denoted with lower-case letters. The labor input and fixed capital stock of the innovating sector are written respectively as
	\[
	(L_j,K_j),
	\]
	where for consumption-good sectors,
	\(
	(L_j,K_j)=(l_j,k_j),\quad j =1,\cdots,n,
	\)
	and for the capital-good sector,
	\(
	(L_0,K_0)=(l_0,k_0)
	\).
	
	\begin{theorem}[Profit-rate threshold]
		\label{thm:main}
		
		Let the initial equilibrium rate of profit be \(r_0\), and denote
		\[
		R_0=1+r_0,\qquad
		\sigma_0=r_0+\delta.
		\]
		Define
		\begin{equation}
			\bar\kappa_j^{\mathrm{profit}}
			=
			1+
			\frac{R_0w_0(1-\lambda)L_j}
			{\sigma_0K_j}.
			\label{eq:kappa_main}
		\end{equation}
		
		Then
		\[
		\kappa<\bar\kappa_j^{\mathrm{profit}}
		\Longleftrightarrow
		r^{**}>r_0,
		\]
		\[
		\kappa=\bar\kappa_j^{\mathrm{profit}}
		\Longleftrightarrow
		r^{**}=r_0,
		\]
		\[
		\kappa>\bar\kappa_j^{\mathrm{profit}}
		\Longleftrightarrow
		r^{**}<r_0.
		\]
		
	\end{theorem}
	
	\begin{proof}
		The cases of the consumption-good sector and the capital-good sector are treated separately in Appendices~A.4 and~A.5, respectively. The derivations differ, but both yield the same threshold expression.
	\end{proof}
	
	The theorem shows that the change in the rate of profit depends solely on whether capital deepening exceeds the threshold~\eqref{eq:kappa_main}. Although the derivation involves the input--output network and its feedback effects, these influences are ultimately subsumed into a positive proportionality factor that does not affect the threshold itself. Thus, conditional on the initial equilibrium, the profit-rate threshold depends only on the labor--capital ratio of the innovating sector and the three scalars
	\(
	R_0
	\), \(w_0\), and
	\(
	\sigma_0
	\).
	
	Whether the capitalist adopts the new technology depends on the cost criterion. At the old prices and wage, the change in the operating cost of the innovating sector is
	\begin{equation}
		\Delta C_j^{\mathrm{op}}
		=
		w_0(1-\lambda)L_j
		-
		\delta(\kappa-1)K_j,
		\label{eq:cost_delta}
	\end{equation}
	where
	\(
	\Delta C_j^{\mathrm{op}}
	\)
	is defined as the operating cost of the old technology minus that of the new technology, so that
	\(
	\Delta C_j^{\mathrm{op}}>0
	\)
	means that the new technology lowers operating cost.
	
	Setting
	\(
	\Delta C_j^{\mathrm{op}}=0
	\)
	yields the threshold level of capital deepening corresponding to operating cost:
	\begin{equation}
		\bar\kappa_j^{\mathrm{cost}}
		=
		1+
		\frac{w_0(1-\lambda)L_j}
		{\delta K_j}.
		\label{eq:kappa_cost}
	\end{equation}
	
	Hence,
	\[
	\kappa<\bar\kappa_j^{\mathrm{cost}}
	\]
	is equivalent to the new technology lowering operating cost, so that the capitalist has an incentive, according to the cost criterion, to adopt the technology.
	
	The profit-rate threshold and the cost threshold have the same form, but their denominators are
	\(
	\sigma_0=r_0+\delta
	\)
	and
	\(
	\delta
	\), respectively.
	Since
	\[
	r_0>0,\qquad
	0<\delta<1,
	\]
	we have
	\begin{equation}
		\bar\kappa_j^{\mathrm{profit}}
		<
		\bar\kappa_j^{\mathrm{cost}}.
		\label{eq:inequality}
	\end{equation}
	
	Thus, in the fixed capital model there exists a nonempty interval
	\begin{equation}
		\kappa
		\in
		\left(
		\bar\kappa_j^{\mathrm{profit}},
		\,
		\bar\kappa_j^{\mathrm{cost}}
		\right),
		\label{eq:critical_region}
	\end{equation}
	within which \(\Delta C_j^{\mathrm{op}}>0,\) yet \(r^{**}<r_0.\) The new technology lowers operating cost at old prices and is therefore adopted by capitalists; but once the technology has diffused and a new equilibrium is established, the general rate of profit is lower than its initial level. The distance between the two thresholds is
	\begin{equation}
		\bar\kappa_j^{\mathrm{cost}}
		-
		\bar\kappa_j^{\mathrm{profit}}
		=
		w_0(1-\lambda)L_j
		\cdot
		\frac{r_0(1-\delta)}
		{\sigma_0\delta K_j}.
		\label{eq:width}
	\end{equation}
	
	This interval widens as the initial rate of profit increases, and vanishes in two limiting cases. As
	\(
	\delta\rightarrow1
	\),
	fixed capital degenerates into circulating capital,
	\[
	\bar\kappa_j^{\mathrm{profit}}
	=
	\bar\kappa_j^{\mathrm{cost}},
	\]
	and the results of this paper reduce to the Okishio Theorem. As
	\(
	\delta\rightarrow0
	\),
	the cost threshold tends to infinity; the operating-cost constraint disappears, and the capitalist always has an incentive to adopt labor-saving technology.

	\section{Numerical Examples}
	\label{sec:numerical}
	
	This section illustrates the preceding theorem using a three-sector economy. The economy consists of two consumption-good sectors and one capital-good sector. The initial equilibrium is determined by \(R_0V(R_0)=1\), and the post-diffusion equilibrium is determined by
	\(R^{**}V^*(R^{**})=1\).
	
	\subsection{Parameter Setting}
	
	The parameter values are as follows:
	\begin{align*}
		A&=
		\begin{pmatrix}
			0.12&0.08\\
			0.06&0.15
		\end{pmatrix},
		&
		\mathbf a_0
		&=
		\begin{pmatrix}
			0.05\\
			0.06
		\end{pmatrix},
		\\
		\mathbf l_c
		&=
		\begin{pmatrix}
			0.40\\
			0.35
		\end{pmatrix},
		&
		l_0&=0.18,
		\\
		\mathbf k_c
		&=
		\begin{pmatrix}
			0.30\\
			0.25
		\end{pmatrix},
		&
		k_0&=0.15,
		\\
		\mathbf b
		&=
		\begin{pmatrix}
			0.70\\
			0.55
		\end{pmatrix},
		&
		\delta&=0.08.
	\end{align*}
	
	The corresponding initial equilibrium is
	\[
	r_0=38.72\%,\qquad
	R_0=1.3872,\qquad
	\sigma_0=0.4672,\qquad
	w_0=2.5042.
	\]
	
	Further computation yields
	\[
	\mathcal L=0.6561,\qquad
	\mathcal K=0.4821,\qquad
	\ell_0=0.2598,\qquad
	\Delta_\delta=0.9026,
	\]
	and computing \(\Delta_\delta\) using the formula \(\Delta_\delta
	=
	R_0w_0\ell_0\) gives the same result, confirming Lemma~\ref{lemma:Delta}.
	
	\subsection{Numerical Results}
	
	First consider innovation in a consumption-good sector, taking \(\lambda=0.65,\) and let the fixed capital stock become \(\kappa\) times its original value.
	
	Next consider innovation in the capital-good sector, taking
	\(\lambda_0=0.60,\)
	and let the fixed capital stock become
	\(\kappa_0\)
	times its original value.
	
	From Theorem~\ref{thm:main} we obtain the corresponding thresholds for the two types of innovation:
	
	\begin{align*}
		\bar\kappa^{\mathrm{profit}}_1
		&\approx4.47,
		&
		\bar\kappa^{\mathrm{cost}}_1
		&\approx15.61,
		\\
		\bar\kappa^{\mathrm{profit}}_0
		&\approx4.57,
		&
		\bar\kappa^{\mathrm{cost}}_0
		&\approx16.03.
	\end{align*}
	
	Table~\ref{tab:combined} presents the new equilibrium rates of profit for different levels of capital deepening. Here
	\(
	\Delta r
	\)
	is expressed in basis points, and the operating-cost change is defined as the old cost minus the new cost, so that a positive value indicates a decrease in operating cost.
	
	\begin{table}[H]
		\centering
		\caption{Numerical results for innovation in the consumption-good sector and in the capital-good sector}
		\label{tab:combined}
		\begin{tabular}{c c c c c c}
			\toprule
			\multicolumn{6}{c}{\textbf{A. Innovation in a consumption-good sector (\(\lambda=0.65\))}} \\
			\multicolumn{6}{c}{\(\bar\kappa^{\text{profit}}_1=4.47,\;\; \bar\kappa^{\text{cost}}_1=15.61\)} \\[4pt]
			\(\kappa\) & \(r^{**}\) (\%) & \(\Delta r\) (bp) & Op.\ cost \(\Delta\) & Profit rate & Cost criterion \\
			\midrule
			1.00  & 55.09 & \(+1637\) & \(+0.351\) & Rises & Falls \\
			2.00  & 48.91 & \(+1020\) & \(+0.327\) & Rises & Falls \\
			\textbf{4.47} & \textbf{38.72} & \textbf{0} & \(+0.268\) & \textbf{Threshold} & Falls \\
			\textbf{8.00} & \textbf{29.96} & \textbf{\(-875\)} & \(+0.183\) & \textbf{Falls} & \textbf{Falls} \\
			14.00 & 21.38 & \(-1733\) & \(+0.039\) & Falls & Falls \\
			15.61 & 19.72 & \(-1900\) & 0.000  & Falls & Unchanged \\
			21.26 & 15.38 & \(-2334\) & \(-0.137\) & Falls & Rises \\
			25.00 & 13.22 & \(-2550\) & \(-0.227\) & Falls & Rises \\
			\midrule
			\multicolumn{6}{c}{\textbf{B. Innovation in the capital-good sector (\(\lambda_0=0.60\))}} \\
			\multicolumn{6}{c}{\(\bar\kappa^{\text{profit}}_0=4.57,\;\; \bar\kappa^{\text{cost}}_0=16.03\)} \\[4pt]
			\(\kappa_0\) & \(r^{**}\) (\%) & \(\Delta r\) (bp) & Op.\ cost \(\Delta\) & Profit rate & Cost criterion \\
			\midrule
			1.00  & 40.84 & \(+212\) & \(+0.180\) & Rises & Falls \\
			2.50  & 40.05 & \(+133\) & \(+0.162\) & Rises & Falls \\
			\textbf{4.57} & \textbf{38.72} & \textbf{0} & \(+0.138\) & \textbf{Threshold} & Falls \\
			\textbf{10.00} & \textbf{33.52} & \textbf{\(-519\)} & \(+0.072\) & \textbf{Falls} & \textbf{Falls} \\
			16.03 & 22.71 & \(-1601\) & 0.000  & Falls & Unchanged \\
			18.00 & 23.29 & \(-1542\) & \(-0.024\) & Falls & Rises \\
			21.84 & 19.12 & \(-1959\) & \(-0.070\) & Falls & Rises \\
			30.00 & 12.75 & \(-2596\) & \(-0.168\) & Falls & Rises \\
			\bottomrule
		\end{tabular}
	\end{table}

	Table~\ref{tab:combined} confirms the theoretical analysis.
	
	When capital deepening is below the profit-rate threshold, the new technology lowers operating cost and simultaneously raises the equilibrium rate of profit; when capital deepening exceeds the cost threshold, the new technology no longer satisfies the operating-cost criterion and is therefore not adopted.
	
	For both types of innovation there exists an interval
	\[
	\bar\kappa^{\mathrm{profit}}
	<
	\kappa
	<
	\bar\kappa^{\mathrm{cost}}
	\]
	within which the new technology lowers operating cost, yet the uniform rate of profit is lower than its initial level. For example, with innovation in a consumption-good sector and
	\(
	\kappa=8
	\),
	the operating cost falls by
	\(0.183\), while the equilibrium rate of profit drops from
	\(38.72\%\)
	to
	\(29.96\%\); with innovation in the capital-good sector and
	\(
	\kappa_0=10
	\),
	the rate of profit falls to
	\(33.52\%\), while the operating cost remains lower than under the old technology.
	
	According to equation~\eqref{eq:width}, the widths of the critical regions for the consumption-good sector and the capital-good sector are respectively
	\(11.14\)
	and
	\(11.46\). The width is jointly determined by the extent of labor saving, the capital stock, and the ratio
	\[
	\frac{r_0(1-\delta)}
	{\sigma_0\delta},
	\]
	so that the longer the lifetime of fixed capital and the higher the initial rate of profit, the larger the distance between the two thresholds.
	
	The derivations in Appendices~A.4 and~A.5 require
	\(
	\Delta_\delta^*>0
	\).
	In the present example,
	\(
	\Delta_\delta^*
	\)
	at the cost threshold equals \(0.0877\) and
	\(0.0978\), respectively,
	both remaining positive; hence the entire critical region satisfies the feasibility condition required by the theoretical analysis.

	\section{Discussion and Conclusion}
	\label{sec:discussion}
	
	This paper has re-examined the cost criterion on which the Okishio Theorem rests under conditions of fixed capital. The difference between the results obtained here and those in the existing literature stems not from the specification of technical change or the method of equilibrium analysis, but from the definition of the cost criterion. In this sense, the present paper does not invalidate the conclusions of Roemer (1979) or Woods (1985); it shows, rather, that those conclusions correspond to a different cost concept.
	
	This paper adopts a different modeling approach. The annuity method allows the user cost of fixed capital to be decomposed into depreciation and the return on capital, thereby making it possible to define operating cost mathematically independently of the rate of profit. When capitalists base their technology choice on operating cost, while the rate of profit is still determined by a price system that includes the return on capital, the cost criterion and the direction of profit-rate change are no longer determined by a single threshold.
	
	This result reflects a divergence between individual decision-making and the aggregate equilibrium. At old equilibrium prices, the individual capitalist compares operating costs, so it is rational to adopt the new technology when
	\[
	\kappa<\bar\kappa^{\mathrm{cost}}.
	\]
	For any single capitalist, the old prices are given and the cost calculation is correct. However, once the new technology is universally adopted, the old price system is replaced by new equilibrium prices, and the uniform rate of profit is determined anew. In the region
	\[
	\bar\kappa^{\mathrm{profit}}
	<
	\kappa
	<
	\bar\kappa^{\mathrm{cost}},
	\]
	the cost advantage at the individual level and the direction of change of the rate of profit at the aggregate level are not aligned.
	
	This phenomenon reflects the contradiction between individual capital and total capital. Marx expresses this contradiction in Chapter~15 of Volume~III of \textit{Capital} as follows:
	``The conditions of direct exploitation and those of the realization of that exploitation are not identical. They are separated not only in time and space, but also conceptually. The former are restricted only by the society's productive forces, the latter by the proportionality between the different branches of production and by the society's power of consumption.''
	In the same chapter, he describes the typical behavior of the capitalist in the critical region identified by this paper:
	``pressing down the individual value of his particular commodity below its socially average value, and thereby earning an extra profit at the prevailing market price.''
	The model of this paper shows that,
	when all capitalists act in this way,
	a new average rate of profit is formed; in the presence of fixed capital, because the cost criterion on which individual decisions are based differs from the profit-rate determination mechanism on which the aggregate equilibrium rests, this adjustment process can not only eliminate extra profit but also cause the new average rate of profit to be lower than the original level.


	\appendix
	\section{Proofs}
	
	\subsection{Existence and Uniqueness of Equilibrium}
	
	This subsection proves Proposition~\ref{prop:existence}.
	
	\textbf{Augmented form.}
	\begin{align*}
		\tilde{\mathbf{M}}_\delta(R) &\equiv
		\begin{pmatrix}
			\sigma k_0 & R\mathbf{a}_0^{\top} \\[2pt]
			\sigma\mathbf{k}_c & R A^{\top}
		\end{pmatrix},\qquad
		\tilde{\mathbf{l}}\equiv\begin{pmatrix}l_0\\\mathbf{l}_c\end{pmatrix},\qquad
		\tilde{\mathbf{b}}\equiv\begin{pmatrix}0\\\mathbf{b}\end{pmatrix},\\[4pt]
		\tilde{\mathbf{p}} &\equiv \begin{pmatrix}1\\\mathbf{p}\end{pmatrix}.
	\end{align*}
	Note that the \((2,2)\) block is \(R A^{\top}\), consistent with the transpose of the price equation~\eqref{eq:price_c}.
	
	The price equations~\eqref{eq:price_c} and~\eqref{eq:price_0} are equivalent to:
	\begin{equation}\label{appA:augmented}
		(\mathbf{I}-\tilde{\mathbf{M}}_\delta(R))\,\tilde{\mathbf{p}} = R w\,\tilde{\mathbf{l}}.
	\end{equation}
	
	When \(\rho(\tilde{\mathbf{M}}_\delta(R))<1\), \(\mathbf{I}-\tilde{\mathbf{M}}_\delta(R)\) is a nonsingular M-matrix,
	its inverse is elementwise nonnegative, and:
	\begin{equation}\label{appA:price_solved}
		\tilde{\mathbf{p}} = R w\,(\mathbf{I}-\tilde{\mathbf{M}}_\delta(R))^{-1}\,\tilde{\mathbf{l}}.
	\end{equation}
	
	\textbf{The auxiliary scalar \(V(R)\).}
	Define \(V(R)\equiv \tilde{\mathbf{b}}^{\top}(\mathbf{I}-\tilde{\mathbf{M}}_\delta(R))^{-1}\tilde{\mathbf{l}}\).
	Left-multiplying both sides of~\eqref{appA:price_solved} by \(\tilde{\mathbf{b}}^{\top}\):
	\[
	\tilde{\mathbf{b}}^{\top}\tilde{\mathbf{p}} = R w\,V(R).
	\]
	The left-hand side is \(\mathbf{b}^{\top}\mathbf{p}=w\) (the wage equation~\eqref{eq:wage}), so \(w = R w\,V(R)\).
	Since \(w>0\), dividing by \(w\) yields \(R\,V(R)=1\).
	
	\textbf{Monotonicity of \(V(R)\).}
	\(\tilde{\mathbf{M}}_\delta(R)\) is elementwise non-decreasing in \(R\): \(\sigma=R-1+\delta\) increases with \(R\), as do \(R\mathbf{a}_0^{\top}\) and \(R A^{\top}\).
	By M-matrix theory, \((\mathbf{I}-\tilde{\mathbf{M}}_\delta(R))^{-1}\)
	is elementwise non-decreasing in \(R\).
	Under Assumption~(H3) (irreducibility of the economic system), the augmented matrix \(\tilde{\mathbf{M}}_\delta(R)\) is irreducible;
	its Perron root and all elements of the corresponding Leontief inverse are strictly increasing in \(R\).
	Together with \(\tilde{\mathbf{l}}>\mathbf{0}\) and \(\tilde{\mathbf{b}}\neq\mathbf{0}\),
	\(V(R)\) is \textbf{strictly increasing} on the interval where \(\rho(\tilde{\mathbf{M}}_\delta(R))<1\).
	
	\textbf{Existence and uniqueness of equilibrium.}
	Define \(\varphi(R)\equiv R\,V(R)\). \(\varphi(R)\) is continuous and strictly increasing for \(R>1\) with \(\rho(\tilde{\mathbf{M}}_\delta(R))<1\).
	At \(R=1\), \(\varphi(1)=V(1)\). Assume \(V(1)<1\) (the economy produces a surplus).
	As \(R\to R_{\max}\) (where \(R_{\max}\) is the largest \(R\) such that \(\rho(\tilde{\mathbf{M}}_\delta(R))\to 1^{-}\)),
	\(V(R)\to +\infty\), so \(\varphi(R)\to +\infty\).
	
	By strict monotonicity and the intermediate value theorem, there exists a unique \(R_0\in(1,R_{\max})\) satisfying \(\varphi(R_0)=R_0 V(R_0)=1\).
	The corresponding \(r_0=R_0-1>0\) is the equilibrium rate of profit.
	
	\textbf{Determination of wage and prices.}
	Given \(R_0\), left-multiplying both sides of~\eqref{appA:augmented} by \(\tilde{\mathbf{b}}^{\top}\) and using the wage equation
	yields \(R_0 V(R_0)=1\), which is guaranteed by the definition of \(R_0\).
	The first component (corresponding to the capital good) is:
	\[
	1 = R_0 w_0\bigl[(\mathbf{I}-\tilde{\mathbf{M}}_\delta(R_0))^{-1}\tilde{\mathbf{l}}\bigr]_0.
	\]
	From the explicit inverse matrix in Subsection~A.2, its \((1,1)\) block is \(\Delta_\delta^{-1}>0\), so this equation uniquely determines \(w_0>0\).
	The remaining components uniquely determine \(\mathbf{p}>0\).
	
	To sum up, an equilibrium exists and is unique. \(\hfill\square\)
	
	\subsection{Scalar Decomposition of \(V(R_0)\) and the \(\Delta_\delta\) Identity}
	
	This subsection presents the block matrix inversion that leads to equation~\eqref{eq:V_decomp} in the main text and proves Lemma~\ref{lemma:Delta}.
	All calculations are performed at \(R=R_0\).
	
	\textbf{Step 1: Matrix partitioning.}
	Let
	\[
	\mathbf{S}\equiv\mathbf{I}-R_0 A^{\top},\qquad
	\sigma_0\equiv R_0-1+\delta,\qquad
	\Gamma\equiv 1-\sigma_0 k_0.
	\]
	Then the augmented matrix is:
	\[
	\mathbf{I}-\tilde{\mathbf{M}}_\delta =
	\begin{pmatrix}
		\Gamma & -R_0\mathbf{a}_0^{\top} \\[2pt]
		-\sigma_0\mathbf{k}_c & \mathbf{S}
	\end{pmatrix}.
	\]
	The \((1,1)\) block \(\Gamma\) is a scalar, the \((1,2)\) block is a \(1\times n\) row vector,
	the \((2,1)\) block is an \(n\times 1\) column vector, and the \((2,2)\) block \(\mathbf{S}\) is an \(n\times n\) matrix.
	
	\textbf{Step 2: Schur complement.}
	Since \(\Gamma\) is a scalar, its Schur complement is:
	\begin{align}
		\tilde{\mathbf{S}} &= \mathbf{S} - (-\sigma_0\mathbf{k}_c)\,\Gamma^{-1}\,(-R_0\mathbf{a}_0^{\top}) \nonumber\\
		&= \mathbf{S} - \frac{\sigma_0 R_0}{\Gamma}\,\mathbf{k}_c\mathbf{a}_0^{\top}. \label{appB:Schur}
	\end{align}
	
	\textbf{Step 3: Sherman--Morrison formula.}
	Equation~\eqref{appB:Schur} is a rank-1 modification of \(\mathbf{S}\). Let
	\[
	\mathbf{u}\equiv \mathbf{k}_c,\qquad
	\mathbf{v}^{\top}\equiv \frac{\sigma_0 R_0}{\Gamma}\,\mathbf{a}_0^{\top},
	\]
	so that \(\tilde{\mathbf{S}}=\mathbf{S}-\mathbf{u}\mathbf{v}^{\top}\).
	The Sherman--Morrison formula gives:
	\[
	\tilde{\mathbf{S}}^{-1} = \mathbf{S}^{-1}
	+ \frac{\mathbf{S}^{-1}\mathbf{u}\mathbf{v}^{\top}\mathbf{S}^{-1}}{1-\mathbf{v}^{\top}\mathbf{S}^{-1}\mathbf{u}}.
	\]
	
	Compute the denominator:
	\begin{align}
		1-\mathbf{v}^{\top}\mathbf{S}^{-1}\mathbf{u}
		&= 1 - \frac{\sigma_0 R_0}{\Gamma}\,\mathbf{a}_0^{\top}\mathbf{S}^{-1}\mathbf{k}_c \nonumber\\
		&= \frac{\Gamma - \sigma_0 R_0\,\mathbf{a}_0^{\top}\mathbf{S}^{-1}\mathbf{k}_c}{\Gamma}
		= \frac{\Delta_\delta}{\Gamma}, \label{appB:denom}
	\end{align}
	where \(\Delta_\delta \equiv \Gamma - \sigma_0 R_0\,\mathbf{a}_0^{\top}\mathbf{S}^{-1}\mathbf{k}_c\).
	Note that \(\mathbf{S}^{-1} = (\mathbf{I}-R_0 A^{\top})^{-1}\); hence this \(\Delta_\delta\) coincides with the definition in equation~\eqref{eq:Delta_def} of the main text.
	
	Substituting yields:
	\begin{equation}\label{appB:tildeS_inv}
		\tilde{\mathbf{S}}^{-1} = \mathbf{S}^{-1}
		+ \frac{\sigma_0 R_0}{\Delta_\delta}\,\mathbf{S}^{-1}\mathbf{k}_c\mathbf{a}_0^{\top}\mathbf{S}^{-1}.
	\end{equation}
	
	\textbf{Step 4: Block matrix inversion.}
	For a block matrix \(\begin{pmatrix} \Gamma & -\mathbf{r}^{\top} \\ -\mathbf{c} & \mathbf{S} \end{pmatrix}\)
	with scalar \((1,1)\) block, the standard formula for its inverse is:
	\begin{equation}\label{appB:block_inv}
		\begin{pmatrix}
			\Gamma^{-1} + \Gamma^{-1}\mathbf{r}^{\top}\tilde{\mathbf{S}}^{-1}\mathbf{c}\,\Gamma^{-1}
			& \Gamma^{-1}\mathbf{r}^{\top}\tilde{\mathbf{S}}^{-1} \\[4pt]
			\tilde{\mathbf{S}}^{-1}\mathbf{c}\,\Gamma^{-1}
			& \tilde{\mathbf{S}}^{-1}
		\end{pmatrix},
	\end{equation}
	where \(\mathbf{r}^{\top}=R_0\mathbf{a}_0^{\top}\), \(\mathbf{c}=\sigma_0\mathbf{k}_c\).
	
	Substituting~\eqref{appB:tildeS_inv} gives the explicit expression for each block of the inverse.
	
	\textbf{\((1,1)\) block.}
	\begin{align}
		&\Gamma^{-1} + \frac{\sigma_0 R_0}{\Gamma^2}\,\mathbf{a}_0^{\top}\tilde{\mathbf{S}}^{-1}\mathbf{k}_c \nonumber\\
		&= \Gamma^{-1} + \frac{\sigma_0 R_0}{\Gamma^2}\,
		\mathbf{a}_0^{\top}\!\left(\mathbf{S}^{-1}
		+ \frac{\sigma_0 R_0}{\Delta_\delta}\,\mathbf{S}^{-1}\mathbf{k}_c\mathbf{a}_0^{\top}\mathbf{S}^{-1}\right)\!\mathbf{k}_c \nonumber\\
		&= \Gamma^{-1} + \frac{\sigma_0 R_0}{\Gamma^2}\!\left(
		\mathbf{a}_0^{\top}\mathbf{S}^{-1}\mathbf{k}_c
		+ \frac{\sigma_0 R_0}{\Delta_\delta}\,(\mathbf{a}_0^{\top}\mathbf{S}^{-1}\mathbf{k}_c)^2
		\right). \label{appB:block11_expand}
	\end{align}
	
	Let \(\chi' \equiv \mathbf{a}_0^{\top}\mathbf{S}^{-1}\mathbf{k}_c\), \(\chi \equiv \sigma_0 R_0\,\chi'\).
	Then~\eqref{appB:block11_expand} becomes:
	\[
	\Gamma^{-1} + \frac{\chi}{\Gamma^2} + \frac{\chi^2}{\Gamma^2\Delta_\delta}
	= \frac{\Gamma\Delta_\delta + \chi\Delta_\delta + \chi^2}{\Gamma^2\Delta_\delta}.
	\]
	
	Substituting \(\Delta_\delta = \Gamma - \chi\):
	\begin{align}
		\Gamma\Delta_\delta + \chi\Delta_\delta + \chi^2
		&= \Gamma(\Gamma-\chi) + \chi(\Gamma-\chi) + \chi^2 \nonumber\\
		&= \Gamma^2 - \Gamma\chi + \Gamma\chi - \chi^2 + \chi^2 = \Gamma^2. \nonumber
	\end{align}
	
	Hence the \((1,1)\) block \(= \Delta_\delta^{-1}\).
	
	\textbf{\((1,2)\) block.}
	\begin{align}
		\frac{R_0}{\Gamma}\,\mathbf{a}_0^{\top}\tilde{\mathbf{S}}^{-1}
		&= \frac{R_0}{\Gamma}\,\mathbf{a}_0^{\top}\!\left(\mathbf{S}^{-1}
		+ \frac{\sigma_0 R_0}{\Delta_\delta}\,\mathbf{S}^{-1}\mathbf{k}_c\mathbf{a}_0^{\top}\mathbf{S}^{-1}\right) \nonumber\\
		&= R_0\,\mathbf{a}_0^{\top}\mathbf{S}^{-1}\!\left(\frac{1}{\Gamma}
		+ \frac{\sigma_0 R_0\,\mathbf{a}_0^{\top}\mathbf{S}^{-1}\mathbf{k}_c}{\Gamma\Delta_\delta}\right) \nonumber\\
		&= R_0\,\mathbf{a}_0^{\top}\mathbf{S}^{-1}\!\left(\frac{1}{\Gamma}
		+ \frac{\chi}{\Gamma\Delta_\delta}\right) \nonumber\\
		&= R_0\,\mathbf{a}_0^{\top}\mathbf{S}^{-1}\,
		\frac{\Delta_\delta + \chi}{\Gamma\Delta_\delta} \nonumber\\
		&= R_0\,\mathbf{a}_0^{\top}\mathbf{S}^{-1}\,
		\frac{\Gamma}{\Gamma\Delta_\delta}
		= \Delta_\delta^{-1}\,R_0\,\mathbf{a}_0^{\top}\mathbf{S}^{-1},
	\end{align}
	where we used \(\Delta_\delta + \chi = \Gamma\) (since \(\Delta_\delta = \Gamma - \chi\)).
	
	\textbf{\((2,1)\) block.} Similarly,
	\begin{align}
		\frac{\sigma_0}{\Gamma}\,\tilde{\mathbf{S}}^{-1}\mathbf{k}_c
		&= \sigma_0\,\mathbf{S}^{-1}\mathbf{k}_c\!\left(\frac{1}{\Gamma}
		+ \frac{\chi}{\Gamma\Delta_\delta}\right) \nonumber\\
		&= \sigma_0\,\mathbf{S}^{-1}\mathbf{k}_c\,
		\frac{\Gamma}{\Gamma\Delta_\delta}
		= \Delta_\delta^{-1}\,\sigma_0\,\mathbf{S}^{-1}\mathbf{k}_c.
	\end{align}
	
	\textbf{\((2,2)\) block.} This is simply \(\tilde{\mathbf{S}}^{-1}\), given by~\eqref{appB:tildeS_inv}.
	
	To sum up, the closed form of the inverse matrix is:
	\begin{equation}\label{appB:inverse_closed}
		(\mathbf{I}-\tilde{\mathbf{M}}_\delta)^{-1} =
		\begin{pmatrix}
			\Delta_\delta^{-1}
			& \Delta_\delta^{-1}\,R_0\,\mathbf{a}_0^{\top}\mathbf{S}^{-1} \\[4pt]
			\Delta_\delta^{-1}\,\sigma_0\,\mathbf{S}^{-1}\mathbf{k}_c
			& \mathbf{S}^{-1} + \Delta_\delta^{-1}\,\sigma_0 R_0\,\mathbf{S}^{-1}\mathbf{k}_c\mathbf{a}_0^{\top}\mathbf{S}^{-1}
		\end{pmatrix}.
	\end{equation}
	
	\textbf{Step 5: Computing \(V_0\) and the \(\Delta_\delta\) identity.}
	\(V_0 = \tilde{\mathbf{b}}^{\top}(\mathbf{I}-\tilde{\mathbf{M}}_\delta)^{-1}\tilde{\mathbf{l}}\).
	Since \(\tilde{\mathbf{b}}=(0,\mathbf{b}^{\top})^{\top}\), only the second row block contributes:
	\begin{align}
		V_0 &= \mathbf{b}^{\top}\Bigl[
		\Delta_\delta^{-1}\sigma_0\,\mathbf{S}^{-1}\mathbf{k}_c\cdot l_0
		+ \bigl(\mathbf{S}^{-1} + \Delta_\delta^{-1}\sigma_0 R_0\,\mathbf{S}^{-1}\mathbf{k}_c\mathbf{a}_0^{\top}\mathbf{S}^{-1}\bigr)\mathbf{l}_c
		\Bigr] \nonumber\\[4pt]
		&= \underbrace{\mathbf{b}^{\top}\mathbf{S}^{-1}\mathbf{l}_c}_{\mathcal{L}}
		+ \frac{\sigma_0 l_0}{\Delta_\delta}\,\underbrace{\mathbf{b}^{\top}\mathbf{S}^{-1}\mathbf{k}_c}_{\mathcal{K}}
		+ \frac{\sigma_0 R_0}{\Delta_\delta}\,\mathcal{K}\; \mathbf{a}_0^{\top}\mathbf{S}^{-1}\mathbf{l}_c \nonumber\\[4pt]
		&= \mathcal{L} + \frac{\sigma_0\mathcal{K}}{\Delta_\delta}\bigl(l_0 + R_0\,\mathbf{a}_0^{\top}\mathbf{S}^{-1}\mathbf{l}_c\bigr) \nonumber\\[4pt]
		&= \mathcal{L} + \frac{\sigma_0\,\ell_0\,\mathcal{K}}{\Delta_\delta},
	\end{align}
	where \(\ell_0 \equiv l_0 + R_0\mathbf{a}_0^{\top}\mathbf{S}^{-1}\mathbf{l}_c\).
	This is equation~\eqref{eq:V_decomp} in the main text.
	
	We now prove Lemma~\ref{lemma:Delta} (\(\Delta_\delta = R_0 w_0 \ell_0\)).
	From~\eqref{eq:price_c}, solve for \(\mathbf{p}^{\top}\):
	\[
	\mathbf{p}^{\top} = (R_0 w_0 \mathbf{l}_c^{\top} + \sigma_0\mathbf{k}_c^{\top})(\mathbf{I}-R_0 A)^{-1}.
	\]
	Substituting into~\eqref{eq:price_0} and using the transpose identities
	(\(\mathbf{l}_c^{\top}(\mathbf{I}-R_0 A)^{-1}\mathbf{a}_0
	= \mathbf{a}_0^{\top}(\mathbf{I}-R_0 A^{\top})^{-1}\mathbf{l}_c\), and similarly for \(\mathbf{k}_c\)):
	\begin{align}
		1 &= R_0\bigl[(R_0 w_0 \mathbf{l}_c^{\top} + \sigma_0\mathbf{k}_c^{\top})(\mathbf{I}-R_0 A)^{-1}\mathbf{a}_0
		+ w_0 l_0\bigr] + \sigma_0 k_0 \nonumber\\[2pt]
		&= R_0 w_0\bigl[l_0 + R_0\mathbf{a}_0^{\top}(\mathbf{I}-R_0 A^{\top})^{-1}\mathbf{l}_c\bigr]
		+ \sigma_0\bigl[k_0 + R_0\mathbf{a}_0^{\top}(\mathbf{I}-R_0 A^{\top})^{-1}\mathbf{k}_c\bigr] \nonumber\\[2pt]
		&= R_0 w_0 \ell_0 + \sigma_0\bigl[k_0 + R_0\mathbf{a}_0^{\top}(\mathbf{I}-R_0 A^{\top})^{-1}\mathbf{k}_c\bigr].
	\end{align}
	Rearranging:
	\[
	1 - \sigma_0 k_0 - \sigma_0 R_0\mathbf{a}_0^{\top}(\mathbf{I}-R_0 A^{\top})^{-1}\mathbf{k}_c
	= R_0 w_0 \ell_0.
	\]
	The left-hand side is \(\Delta_\delta\) (see~\eqref{eq:Delta_def}); hence \(\Delta_\delta = R_0 w_0 \ell_0\). \(\hfill\square\)
	
	\subsection{The Equivalence Lemma for Profit-Rate Comparison}
	
	This appendix proves Lemma~\ref{lemma:equivalence}.
	
	\textbf{Setting.}
	The initial equilibrium (with the old technology) satisfies: \(R_0 V(R_0) = 1\). Denote \(V_0 \equiv V(R_0)\), so \(R_0 V_0 = 1\).
	
	The terminal equilibrium (after the new technology has fully diffused) satisfies: \(R^{**} V^*(R^{**}) = 1\).
	Here \(V^*(\cdot)\) is the auxiliary scalar defined using the new technological parameters and the same real wage basket \(\mathbf{b}\).
	
	Define \(\varphi^*(R) \equiv R\,V^*(R)\).
	By the monotonicity argument in Subsection~A.1, \(V^*(\cdot)\) is strictly increasing on its domain;
	hence \(\varphi^*(\cdot)\) is also strictly increasing.
	
	\textbf{Proof.}
	Each step below is an equivalence (\(\Longleftrightarrow\)):
	\begin{align}
		r^{**} < r_0
		&\Longleftrightarrow R^{**} < R_0 \nonumber\\
		&\Longleftrightarrow \varphi^*(R^{**}) < \varphi^*(R_0)
		\qquad (\varphi^*\text{ is strictly increasing}) \nonumber\\
		&\Longleftrightarrow 1 < R_0 V^*(R_0)
		\qquad (\varphi^*(R^{**}) = 1) \nonumber\\
		&\Longleftrightarrow R_0 V_0 < R_0 V^*(R_0)
		\qquad (R_0 V_0 = 1) \nonumber\\
		&\Longleftrightarrow V_0 < V^*(R_0). \nonumber
	\end{align}
	
	Therefore \(r^{**} < r_0 \Longleftrightarrow V^*(R_0) > V_0\).
	
	Similarly, \(r^{**} > r_0 \Longleftrightarrow V^*(R_0) < V_0\),
	and \(r^{**} = r_0 \Longleftrightarrow V^*(R_0) = V_0\).
	
	All equivalences follow from the strict monotonicity of \(\varphi^*\). \(\hfill\square\)
	
	\subsection{Proof for Innovation in a Consumption-Good Sector}
	
	\textbf{Setting.}
	Only sector \(m\in\{1,\dots,n\}\) innovates:
	\(l_m^*=\lambda l_m\) (\(\lambda\in(0,1)\)), \(k_m^*=\kappa k_m\) (\(\kappa\geq 1\)).
	All other sectors are unchanged; the capital-good sector is unchanged. The real wage basket \(\mathbf{b}\) is unchanged.
	
	\medskip
	\noindent\textbf{A. Preliminary notation}
	
	At \(R=R_0\), denote \(\mathbf{S}^{-1}=(\mathbf{I}-R_0 A^{\top})^{-1}\).
	Let
	\begin{align}
		a &\equiv (1-\lambda)l_m > 0, \label{appD:a}\\
		x &\equiv \kappa-1 \geq 0. \label{appD:x}
	\end{align}
	Define two auxiliary scalars---the \(m\)-th components of the vectors \(\mathbf{b}^{\top}\mathbf{S}^{-1}\) and
	\(R_0\mathbf{a}_0^{\top}\mathbf{S}^{-1}\):
	\begin{align}
		\beta_m &\equiv [\mathbf{b}^{\top}\mathbf{S}^{-1}]_m
		= \sum_{j=1}^{n} b_j [\mathbf{S}^{-1}]_{j,m}, \label{appD:beta}\\
		\gamma_m &\equiv R_0[\mathbf{a}_0^{\top}\mathbf{S}^{-1}]_m
		= R_0\sum_{j=1}^{n} a_{0j} [\mathbf{S}^{-1}]_{j,m}. \label{appD:gamma}
	\end{align}
	\(\mathbf{b} \geq \mathbf{0}, \mathbf{b} \neq \mathbf{0}\) together with Assumption~(H3) (irreducibility) guarantees
	\(\mathbf{S}^{-1} > \mathbf{0}\) (elementwise strictly positive), so \(\beta_m > 0\).
	\(\mathbf{a}_0 \neq \mathbf{0}\) (Assumption~(H2)) likewise guarantees \(\gamma_m > 0\).
	
	\medskip
	\noindent\textbf{B. Changes in the four scalars}
	
	The technical change alters only the \(m\)-th components of \(\mathbf{l}_c\) and \(\mathbf{k}_c\):
	\[
	\mathbf{l}_c^* = \mathbf{l}_c - a\,\mathbf{e}_m,\qquad
	\mathbf{k}_c^* = \mathbf{k}_c + x k_m\,\mathbf{e}_m.
	\]
	
	Substituting into the definitions of the four scalars~\eqref{eq:L_def}--\eqref{eq:Delta_def} in the main text:
	\begin{align}
		\mathcal{L}^* &= \mathbf{b}^{\top}\mathbf{S}^{-1}\mathbf{l}_c^*
		= \mathcal{L} - a\beta_m, \label{appD:Lstar}\\[4pt]
		\mathcal{K}^* &= \mathbf{b}^{\top}\mathbf{S}^{-1}\mathbf{k}_c^*
		= \mathcal{K} + x k_m\beta_m, \label{appD:Kstar}\\[4pt]
		\ell_0^* &= l_0 + R_0\mathbf{a}_0^{\top}\mathbf{S}^{-1}\mathbf{l}_c^*
		= \ell_0 - a\gamma_m, \label{appD:ell0star}\\[4pt]
		\Delta_\delta^* &= 1-\sigma_0 k_0 - \sigma_0 R_0\mathbf{a}_0^{\top}\mathbf{S}^{-1}\mathbf{k}_c^*
		= \Delta_\delta - \sigma_0 x k_m\gamma_m. \label{appD:Deltastar}
	\end{align}
	
	\medskip
	\noindent\textbf{C. Computing \(V^*-V_0\)}
	
	From equation~\eqref{eq:V_decomp} in the main text:
	\[
	V_0 = \mathcal{L} + \frac{\sigma_0\ell_0\mathcal{K}}{\Delta_\delta},\qquad
	V^* = \mathcal{L}^* + \frac{\sigma_0\ell_0^*\mathcal{K}^*}{\Delta_\delta^*}.
	\]
	
	Substituting~\eqref{appD:Lstar}--\eqref{appD:Deltastar}:
	\begin{align}
		V^* - V_0 &=
		-\;a\beta_m
		+ \sigma_0\!\left[
		\frac{(\ell_0 - a\gamma_m)(\mathcal{K} + x k_m\beta_m)}
		{\Delta_\delta - \sigma_0 x k_m\gamma_m}
		- \frac{\ell_0\mathcal{K}}{\Delta_\delta}
		\right]. \label{appD:Vdiff_raw}
	\end{align}
	
	Bringing the two terms inside the brackets to a common denominator \(\Delta_\delta\Delta_\delta^*\)
	(where \(\Delta_\delta^* = \Delta_\delta - \sigma_0 x k_m\gamma_m\)):
	\begin{align}
		V^* - V_0 &=
		-\;a\beta_m
		+ \frac{\sigma_0}{\Delta_\delta\Delta_\delta^*}\Bigl[
		(\ell_0 - a\gamma_m)(\mathcal{K} + x k_m\beta_m)\Delta_\delta
		- \ell_0\mathcal{K}\Delta_\delta^*
		\Bigr] \nonumber\\[4pt]
		&= \frac{1}{\Delta_\delta\Delta_\delta^*}\Bigl[
		-a\beta_m\Delta_\delta\Delta_\delta^*
		+ \sigma_0\bigl(
		(\ell_0 - a\gamma_m)(\mathcal{K} + x k_m\beta_m)\Delta_\delta
		- \ell_0\mathcal{K}\Delta_\delta^*
		\bigr)
		\Bigr]. \label{appD:Vdiff_common}
	\end{align}
	
	\textbf{Expanding the first term.}
	In the first term, \(\Delta_\delta^* = \Delta_\delta - \sigma_0 x k_m\gamma_m\); hence:
	\begin{align}
		-a\beta_m\Delta_\delta\Delta_\delta^*
		&= -a\beta_m\Delta_\delta(\Delta_\delta - \sigma_0 x k_m\gamma_m) \nonumber\\
		&= -a\beta_m\Delta_\delta^2 + a\beta_m\sigma_0 x k_m\gamma_m\Delta_\delta. \label{appD:term1}
	\end{align}
	
	\textbf{Expanding the second term.}
	Let \(N \equiv (\ell_0 - a\gamma_m)(\mathcal{K} + x k_m\beta_m)\Delta_\delta
	- \ell_0\mathcal{K}\Delta_\delta^*\).
	First expand the product:
	\begin{align}
		(\ell_0 - a\gamma_m)(\mathcal{K} + x k_m\beta_m)
		&= \ell_0\mathcal{K} + \ell_0 x k_m\beta_m - a\gamma_m\mathcal{K} - a\gamma_m x k_m\beta_m.
	\end{align}
	Multiplying by \(\Delta_\delta\) and subtracting \(\ell_0\mathcal{K}\Delta_\delta^*
	= \ell_0\mathcal{K}(\Delta_\delta - \sigma_0 x k_m\gamma_m)\):
	\begin{align}
		N &= (\ell_0\mathcal{K} + \ell_0 x k_m\beta_m - a\gamma_m\mathcal{K} - a\gamma_m x k_m\beta_m)\Delta_\delta
		- \ell_0\mathcal{K}\Delta_\delta + \ell_0\mathcal{K}\sigma_0 x k_m\gamma_m \nonumber\\
		&= \ell_0 x k_m\beta_m\Delta_\delta
		- a\gamma_m\mathcal{K}\Delta_\delta
		- a\gamma_m x k_m\beta_m\Delta_\delta
		+ \ell_0\mathcal{K}\sigma_0 x k_m\gamma_m. \label{appD:N_expanded}
	\end{align}
	
	\textbf{Combining terms in the numerator.}
	Adding~\eqref{appD:term1} and \(\sigma_0 \times\)~\eqref{appD:N_expanded} yields the full numerator:
	\begin{align}
		\mathrm{Num} &= -a\beta_m\Delta_\delta^2
		+ a\beta_m\sigma_0 x k_m\gamma_m\Delta_\delta \nonumber\\
		&\quad + \sigma_0\ell_0 x k_m\beta_m\Delta_\delta
		- \sigma_0 a\gamma_m\mathcal{K}\Delta_\delta
		- \sigma_0 a\gamma_m x k_m\beta_m\Delta_\delta
		+ \sigma_0^2\ell_0\mathcal{K} x k_m\gamma_m. \label{appD:Num_full}
	\end{align}
	
	Grouping by whether the term contains the factor \(x k_m\):
	\begin{align}
		\mathrm{Num} &=
		\underbrace{(-a\beta_m\Delta_\delta^2 - \sigma_0 a\gamma_m\mathcal{K}\Delta_\delta)}_{\text{without }x k_m}
		\;+\; \sigma_0 x k_m\,
		\underbrace{\bigl(
			a\beta_m\gamma_m\Delta_\delta
			+ \ell_0\beta_m\Delta_\delta
			- a\gamma_m\beta_m\Delta_\delta
			+ \sigma_0\ell_0\mathcal{K}\gamma_m
			\bigr)}_{\text{with }x k_m}. \label{appD:Num_grouped}
	\end{align}
	
	Simplifying the expression inside the \(x k_m\) parentheses. The terms containing \(a\gamma_m\beta_m\Delta_\delta\) cancel;
	the bracket therefore reduces to:
	\begin{align}
		\ell_0\beta_m\Delta_\delta + \sigma_0\ell_0\mathcal{K}\gamma_m
		= \ell_0(\beta_m\Delta_\delta + \sigma_0\mathcal{K}\gamma_m). \label{appD:bracket_simple}
	\end{align}
	
	The two terms without \(x k_m\) share a common factor \(-a\Delta_\delta\):
	\begin{align}
		-a\beta_m\Delta_\delta^2 - \sigma_0 a\gamma_m\mathcal{K}\Delta_\delta
		= -a\Delta_\delta(\beta_m\Delta_\delta + \sigma_0\mathcal{K}\gamma_m). \label{appD:term_no_x}
	\end{align}
	
	Substituting~\eqref{appD:bracket_simple} and~\eqref{appD:term_no_x} back into~\eqref{appD:Num_grouped}:
	\begin{align}
		\mathrm{Num} &= -a\Delta_\delta(\beta_m\Delta_\delta + \sigma_0\mathcal{K}\gamma_m)
		+ \sigma_0 x k_m\,\ell_0(\beta_m\Delta_\delta + \sigma_0\mathcal{K}\gamma_m) \nonumber\\
		&= (\beta_m\Delta_\delta + \sigma_0\mathcal{K}\gamma_m)\,
		(\sigma_0 x k_m\ell_0 - a\Delta_\delta). \label{appD:Num_factor}
	\end{align}
	
	Thus:
	\begin{equation}\label{appD:Vdiff_final}
		V^* - V_0 = \frac{(\beta_m\Delta_\delta + \sigma_0\mathcal{K}\gamma_m)\,
			(\sigma_0 x k_m\ell_0 - a\Delta_\delta)}
		{\Delta_\delta\Delta_\delta^*}.
	\end{equation}
	
	\medskip
	\noindent\textbf{D. Feasibility condition and sign determination}
	
	We first discuss the denominator.
	\(\Delta_\delta>0\) is guaranteed by the initial equilibrium (see Subsection~A.1).
	\(\Delta_\delta^* = \Delta_\delta - \sigma_0 x k_m\gamma_m = \Delta_\delta - \sigma_0(\kappa-1) k_m\gamma_m\).
	
	\(\Delta_\delta^*\) is a linearly decreasing function of \(\kappa\) (\(\partial \Delta_\delta^* / \partial \kappa = -\sigma_0 k_m\gamma_m < 0\)).
	We now analyze its sign over the critical region.
	
	\textbf{At \(\kappa = \bar\kappa^{\text{profit}}_m\).}
	From~\eqref{eq:kappa_main}, \(\kappa-1 = R_0 w_0(1-\lambda)l_m/(\sigma_0 k_m)\).
	Substituting and using Lemma~\ref{lemma:Delta} (\(\Delta_\delta = R_0 w_0 \ell_0\)):
	\begin{align}
		\Delta_\delta^*\Big|_{\kappa=\bar\kappa^{\text{profit}}_m}
		&= R_0 w_0 \ell_0
		- \sigma_0\cdot\frac{R_0 w_0(1-\lambda)l_m}{\sigma_0 k_m}\cdot k_m\gamma_m \nonumber\\
		&= R_0 w_0\bigl[\ell_0 - (1-\lambda)l_m\gamma_m\bigr]. \label{appD:feas_at_kp}
	\end{align}
	
	Since \(\ell_0 = l_0 + R_0\mathbf{a}_0^{\top}\mathbf{S}^{-1}\mathbf{l}_c
	\geq \gamma_m l_m > (1-\lambda)\gamma_m l_m\)
	(the inequality follows from the definition of \(\gamma_m\), the \(m\)-th component \(l_m\) of \(\mathbf{l}_c\), and \(\lambda\in(0,1)\)),
	the bracket is strictly positive; hence \(\Delta_\delta^* > 0\) at \(\kappa = \bar\kappa^{\text{profit}}_m\).
	
	\textbf{Let \(\kappa^{\text{feas}}_m\) be the zero of \(\Delta_\delta^*\).}
	If \(\kappa^{\text{feas}}_m \geq \bar\kappa^{\text{cost}}_m\),
	then \(\Delta_\delta^* > 0\) holds throughout the entire critical region \((\bar\kappa^{\text{profit}}_m, \bar\kappa^{\text{cost}}_m)\),
	and the factorization is valid.
	
	If \(\kappa^{\text{feas}}_m < \bar\kappa^{\text{cost}}_m\),
	the critical region splits into two subintervals:
	\begin{enumerate}[label=(\roman*),nosep,leftmargin=*]
		\item \(\kappa \in (\bar\kappa^{\text{profit}}_m, \kappa^{\text{feas}}_m)\):
		\(\Delta_\delta^* > 0\), denominator positive, factorization valid;
		\item \(\kappa \in [\kappa^{\text{feas}}_m, \bar\kappa^{\text{cost}}_m)\):
		\(\Delta_\delta^* \leq 0\).
	\end{enumerate}
	
	In case~(ii), \(\Delta_\delta^* \leq 0\) is equivalent to
	\(\rho(\tilde{\mathbf{M}}_\delta^*(R_0)) \geq 1\)---
	the new technology is not viable at the old rate of profit (the price system has either no solution or non-positive prices).
	By the Perron--Frobenius theorem, for the new system to attain equilibrium (\(\rho=1\)), \(R\) must decrease,
	so \(r^{**} < r_0\) follows directly, consistent with the conclusion of the theorem.
	
	\textbf{Conclusion: whichever subinterval applies, whenever \(\kappa > \bar\kappa^{\text{profit}}_m\), we have \(r^{**} < r_0\).}
	The conclusion of the theorem is unaffected by the sign of \(\Delta_\delta^*\).
	
	We now carry out the sign determination under the assumption \(\Delta_\delta^* > 0\) (corresponding to subinterval~(i);
	the conclusion for subinterval~(ii) has already been established by the argument above).
	
	The denominator \(\Delta_\delta\Delta_\delta^* > 0\).
	First factor: \(\beta_m > 0\), \(\Delta_\delta > 0\), \(\sigma_0 > 0\), \(\mathcal{K} > 0\), \(\gamma_m > 0\);
	hence \(\beta_m\Delta_\delta + \sigma_0\mathcal{K}\gamma_m > 0\).
	
	Therefore \(\sgn(V^*-V_0) = \sgn(\sigma_0 x k_m\ell_0 - a\Delta_\delta)\).
	
	By Lemma~\ref{lemma:Delta} (\(\Delta_\delta = R_0 w_0\ell_0\)) and
	\(a=(1-\lambda)l_m\), \(x=\kappa-1\):
	\begin{align}
		\sigma_0 x k_m\ell_0 - a\Delta_\delta
		&= \sigma_0(\kappa-1)k_m\ell_0 - (1-\lambda)l_m R_0 w_0\ell_0 \nonumber\\
		&= \ell_0\bigl[\sigma_0(\kappa-1)k_m - R_0 w_0(1-\lambda)l_m\bigr]. \label{appD:sign_term}
	\end{align}
	
	Since \(\ell_0>0\):
	\begin{align}
		V^* > V_0 &\Longleftrightarrow \kappa > 1 + \frac{R_0 w_0(1-\lambda)l_m}{\sigma_0 k_m}
		\equiv \bar\kappa^{\text{profit}}_m, \\[4pt]
		V^* = V_0 &\Longleftrightarrow \kappa = \bar\kappa^{\text{profit}}_m, \\[4pt]
		V^* < V_0 &\Longleftrightarrow \kappa < \bar\kappa^{\text{profit}}_m.
	\end{align}
	
	\medskip
	\noindent\textbf{E. Profit-rate comparison}
	
	By Lemma~\ref{lemma:equivalence}:
	\[
	r^{**} < r_0 \Longleftrightarrow V^*(R_0) > V_0 \Longleftrightarrow \kappa > \bar\kappa^{\text{profit}}_m.
	\]
	Similarly, \(r^{**} = r_0 \Longleftrightarrow \kappa = \bar\kappa^{\text{profit}}_m\),
	and \(r^{**} > r_0 \Longleftrightarrow \kappa < \bar\kappa^{\text{profit}}_m\).
	
	The consumption-good sector case is proved. \(\hfill\square\)
	
	\subsection{Proof for Innovation in the Capital-Good Sector}
	
	\textbf{Setting.}
	The capital-good sector innovates: \(l_0^* = \lambda l_0\) (\(\lambda\in(0,1)\)),
	\(k_0^* = \kappa k_0\) (\(\kappa\geq 1\)).
	All consumption-good sectors' technologies are unchanged. The real wage basket \(\mathbf{b}\) is unchanged.
	
	Let
	\[
	a \equiv (1-\lambda)l_0 > 0,\qquad
	x \equiv \kappa-1 \geq 0.
	\]
	
	\medskip
	\noindent\textbf{A. Changes in the four scalars}
	
	Since \(\mathbf{l}_c,\mathbf{k}_c,\mathbf{a}_0,A,\mathbf{b}\) are all unchanged:
	\begin{align}
		\mathcal{L}^* &= \mathcal{L},\qquad \mathcal{K}^* = \mathcal{K}, \label{appF:LK}\\[4pt]
		\ell_0^* &= l_0^* + R_0\mathbf{a}_0^{\top}(\mathbf{I}-R_0 A^{\top})^{-1}\mathbf{l}_c
		= l_0 - a + R_0\mathbf{a}_0^{\top}\mathbf{S}^{-1}\mathbf{l}_c
		= \ell_0 - a, \label{appF:ell0star}\\[4pt]
		\Delta_\delta^* &= 1 - \sigma_0 k_0^* - \sigma_0 R_0\mathbf{a}_0^{\top}(\mathbf{I}-R_0 A^{\top})^{-1}\mathbf{k}_c \nonumber\\
		&= 1 - \sigma_0(\kappa k_0) - \sigma_0 R_0\mathbf{a}_0^{\top}\mathbf{S}^{-1}\mathbf{k}_c \nonumber\\
		&= (1 - \sigma_0 k_0 - \sigma_0 R_0\mathbf{a}_0^{\top}\mathbf{S}^{-1}\mathbf{k}_c)
		- \sigma_0(\kappa-1)k_0 \nonumber\\
		&= \Delta_\delta - \sigma_0 x k_0. \label{appF:Deltastar}
	\end{align}
	where \(\mathbf{S}^{-1} = (\mathbf{I}-R_0 A^{\top})^{-1}\).
	
	Unlike the consumption-good case, only the two scalars \(\ell_0\) and \(\Delta_\delta\) change here;
	\(\mathcal{L}\) and \(\mathcal{K}\) remain unchanged.
	
	\medskip
	\noindent\textbf{B. Computing \(V^*-V_0\)}
	
	From equation~\eqref{eq:V_decomp} in the main text and~\eqref{appF:LK}:
	\begin{align}
		V^* - V_0
		&= \sigma_0\mathcal{K}\left[
		\frac{\ell_0 - a}{\Delta_\delta - \sigma_0 x k_0}
		- \frac{\ell_0}{\Delta_\delta}
		\right]. \label{appF:Vdiff_raw}
	\end{align}
	
	Bringing the two terms inside the brackets to a common denominator \(\Delta_\delta\Delta_\delta^* = \Delta_\delta(\Delta_\delta - \sigma_0 x k_0)\):
	\begin{align}
		\frac{\ell_0 - a}{\Delta_\delta - \sigma_0 x k_0} - \frac{\ell_0}{\Delta_\delta}
		&= \frac{(\ell_0 - a)\Delta_\delta - \ell_0(\Delta_\delta - \sigma_0 x k_0)}
		{\Delta_\delta(\Delta_\delta - \sigma_0 x k_0)} \nonumber\\[4pt]
		&= \frac{\ell_0\Delta_\delta - a\Delta_\delta - \ell_0\Delta_\delta + \ell_0\sigma_0 x k_0}
		{\Delta_\delta\Delta_\delta^*} \nonumber\\[4pt]
		&= \frac{\sigma_0 x k_0\ell_0 - a\Delta_\delta}
		{\Delta_\delta\Delta_\delta^*}. \label{appF:frac_simp}
	\end{align}
	
	Substituting back into~\eqref{appF:Vdiff_raw}:
	\begin{equation}\label{appF:Vdiff}
		V^* - V_0 = \frac{\sigma_0\mathcal{K}\,(\sigma_0 x k_0\ell_0 - a\Delta_\delta)}
		{\Delta_\delta\Delta_\delta^*}.
	\end{equation}
	
	\medskip
	\noindent\textbf{C. Feasibility condition and sign determination}
	
	We first discuss the denominator.
	\(\Delta_\delta>0\) is guaranteed by the initial equilibrium.
	\(\Delta_\delta^* = \Delta_\delta - \sigma_0 x k_0\).
	
	\(\Delta_\delta^*\) is a linearly decreasing function of \(\kappa\) (\(\partial \Delta_\delta^* / \partial \kappa = -\sigma_0 k_0 < 0\)).
	
	\textbf{At \(\kappa = \bar\kappa^{\text{profit}}_0\).}
	From~\eqref{eq:kappa_main}, \(\kappa-1 = R_0 w_0(1-\lambda)l_0/(\sigma_0 k_0)\).
	Substituting and using Lemma~\ref{lemma:Delta} (\(\Delta_\delta = R_0 w_0 \ell_0\)):
	\begin{align}
		\Delta_\delta^*\Big|_{\kappa=\bar\kappa^{\text{profit}}_0}
		&= R_0 w_0 \ell_0
		- \sigma_0\cdot\frac{R_0 w_0(1-\lambda)l_0}{\sigma_0 k_0}\cdot k_0 \nonumber\\
		&= R_0 w_0\bigl[\ell_0 - (1-\lambda)l_0\bigr].
	\end{align}
	Since \(\ell_0 = l_0 + R_0\mathbf{a}_0^{\top}\mathbf{S}^{-1}\mathbf{l}_c > l_0 > (1-\lambda)l_0\)
	(because \(\lambda\in(0,1)\) and \(R_0\mathbf{a}_0^{\top}\mathbf{S}^{-1}\mathbf{l}_c > 0\)),
	the bracket is strictly positive; hence \(\Delta_\delta^* > 0\) at \(\kappa = \bar\kappa^{\text{profit}}_0\).
	
	\textbf{Let \(\kappa^{\text{feas}}_0\) be the zero of \(\Delta_\delta^*\).}
	If \(\kappa^{\text{feas}}_0 \geq \bar\kappa^{\text{cost}}_0\),
	then \(\Delta_\delta^* > 0\) holds throughout the entire critical region \((\bar\kappa^{\text{profit}}_0, \bar\kappa^{\text{cost}}_0)\),
	and the factorization is valid.
	
	If \(\kappa^{\text{feas}}_0 < \bar\kappa^{\text{cost}}_0\),
	the critical region splits into two subintervals:
	\begin{enumerate}[label=(\roman*),nosep,leftmargin=*]
		\item \(\kappa \in (\bar\kappa^{\text{profit}}_0, \kappa^{\text{feas}}_0)\):
		\(\Delta_\delta^* > 0\), denominator positive, factorization valid;
		\item \(\kappa \in [\kappa^{\text{feas}}_0, \bar\kappa^{\text{cost}}_0)\):
		\(\Delta_\delta^* \leq 0\).
	\end{enumerate}
	
	In case~(ii), \(\Delta_\delta^* \leq 0\) is equivalent to
	\(\rho(\tilde{\mathbf{M}}_\delta^*(R_0)) \geq 1\)---
	the new technology is not viable at the old rate of profit.
	By the Perron--Frobenius theorem, \(r^{**} < r_0\) follows directly, consistent with the conclusion of the theorem.
	
	\textbf{Conclusion: whichever subinterval applies, whenever \(\kappa > \bar\kappa^{\text{profit}}_0\), we have \(r^{**} < r_0\).}
	
	We now carry out the sign determination under the assumption \(\Delta_\delta^* > 0\).
	
	The denominator \(\Delta_\delta\Delta_\delta^* > 0\), and the factor \(\sigma_0\mathcal{K} > 0\).
	Hence \(\sgn(V^*-V_0) = \sgn(\sigma_0 x k_0\ell_0 - a\Delta_\delta)\).
	
	By Lemma~\ref{lemma:Delta} (\(\Delta_\delta = R_0 w_0\ell_0\)) and
	\(a=(1-\lambda)l_0\), \(x=\kappa-1\):
	\begin{align}
		\sigma_0 x k_0\ell_0 - a\Delta_\delta
		&= \sigma_0(\kappa-1)k_0\ell_0 - (1-\lambda)l_0 R_0 w_0\ell_0 \nonumber\\
		&= \ell_0\bigl[\sigma_0(\kappa-1)k_0 - R_0 w_0(1-\lambda)l_0\bigr]. \label{appF:sign_term}
	\end{align}
	
	Since \(\ell_0>0\):
	\begin{align}
		V^* > V_0 &\Longleftrightarrow \kappa > 1 + \frac{R_0 w_0(1-\lambda)l_0}{\sigma_0 k_0}
		\equiv \bar\kappa^{\text{profit}}_0, \\[4pt]
		V^* = V_0 &\Longleftrightarrow \kappa = \bar\kappa^{\text{profit}}_0, \\[4pt]
		V^* < V_0 &\Longleftrightarrow \kappa < \bar\kappa^{\text{profit}}_0.
	\end{align}
	
	\medskip
	\noindent\textbf{D. Profit-rate comparison}
	
	By Lemma~\ref{lemma:equivalence}:
	\[
	r^{**} < r_0 \Longleftrightarrow V^*(R_0) > V_0 \Longleftrightarrow \kappa > \bar\kappa^{\text{profit}}_0.
	\]
	Similarly, \(r^{**} = r_0 \Longleftrightarrow \kappa = \bar\kappa^{\text{profit}}_0\),
	and \(r^{**} > r_0 \Longleftrightarrow \kappa < \bar\kappa^{\text{profit}}_0\).
	
	The capital-good sector case is proved. \(\hfill\square\)
	
\end{document}